\documentclass[useAMS,usenatbib]{mn2e}
\usepackage{graphicx}
\usepackage{epsfig}
\usepackage{amsmath}
\usepackage{amssymb}
\usepackage{xcolor}
\usepackage{ulem}

\newcommand{\be}{\begin{equation}}
\newcommand{\beq}{\begin{equation}}
\newcommand{\ba}{\begin{eqnarray}}
\newcommand{\ee}{\end{equation}}
\newcommand{\eeq}{\end{equation}}
\newcommand{\ea}{\end{eqnarray}}

\def\lsim{~\rlap{$<$}{\lower 1.0ex\hbox{$\sim$}}}

\def\gsim{~\rlap{$>$}{\lower 1.0ex\hbox{$\sim$}}}

\title[21cm-galaxy cross-power spectrum]{The cross-power spectrum between 21cm emission and galaxies in hierarchical galaxy formation models}	
\author[Jaehong~Park et al.]
      {Jaehong~Park$^{1}$\thanks{jaehongp@student.unimelb.edu.au}, Han-Seek~Kim$^{1}$\thanks{hansikk@unimelb.edu.au}, J. Stuart B.~Wyithe$^{1,3}$, C. G. Lacey$^{2}$\\
       $^1$School of Physics, The University of Melbourne, Parkville, VIC 3010, Australia\\
      $^2$Institute for Computational Cosmology, Department of Physics, University of Durham, South Road, Durham DH1 3LE, UK\\
  	$^3$ARC Centre of Excellence for All-sky Astrophysics (CAASTRO)}
\date{}

\pagerange{\pageref{firstpage}--\pageref{lastpage}}
\pubyear{2013} 

\begin{document}

\maketitle

\label{firstpage}


\begin{abstract}

\noindent 

The correlation between 21cm fluctuations and galaxies is sensitive to the astrophysical properties of the galaxies that drove reionization. Thus, detailed measurements of the cross-power spectrum and its evolution could provide a powerful measurement both of the properties of early galaxies and the process of reionization. In this paper, we study the evolution of the cross-power spectrum between 21cm emission and galaxies using a model which combines the hierarchical galaxy formation model GALFORM  implemented within the Millennium-II dark matter simulation, with a semi-numerical scheme to describe the resulting ionization structure.  We find that inclusion of different feedback processes changes the cross-power spectrum shape and amplitude. In particular, the feature in the cross-power spectrum corresponding to the size of ionized regions is significantly affected by supernovae feedback. We calculate predicted observational uncertainties of the cross-correlation coefficient based on specifications of the Murchison Widefield Array (MWA) combined with galaxy surveys of varying area and depth. We find that the cross-power spectrum could be detected over several square degrees of galaxy survey with galaxy redshift errors $\sigma_{z}\lesssim 0.1$. 

\end{abstract}


\section{Introduction}
 
The prospect of measuring the 21 cm power spectrumÊ from  the  epoch  of  reionization  is  a  focus  of  modern  theoretical  cosmology  \citep[e.g.][]{Morales&Wyithe2010}.   A  very  successful  technique  has  been  to  employ  an ÊN-body  code  to  generate  a  distribution  of  halos,  and  then  applyÊ radiative  transfer  methods  in  post-processingÊ to  model  the  generation  of  ionized  structure  on  large  scales  using  various  models  for  the  ionizing  sources  \citep[e.g.][]{ciardi2003,sokasian2003,iliev2007,zahn2007,trac2007,shin2008,Il08,trac2008}.  However, when constructing  models Êto  assignÊ ionizingÊ luminosities Êto  dark  matter  halos,Ê most  studies  have  used  a  constant  mass-to-luminosity  relation. On  the  other  hand, Êthe  degree to which the important astrophysics governingÊ formation  and  evolution  of  high  redshift  galaxies  will Êinfluence  observations  of  the  21cm  power  spectrum  is  not  well  known.  To  improve  on  the  source  modelling  for  calculation  ofÊ theÊ ionizingÊ photon  budget  in Êreionization  simulations,  several  studies  \citep{Benson2006, Theuns2011, Lacey2011}  have  used  GALFORM Ê\citep{Cole2000,Baugh2005,Bower2006}  combined  withÊ Monte-Carlo  merger  trees.  However  these  studies  calculated  only  theÊ global  evolution  of  reionization,  and  are  not  able  to  address  the  reionization  structure.  Most  recently,  \cite{Kim2012a}  haveÊ combined  GALFORM  implemented  within  the  Millennium-II  dark  matter  simulation  \citep{MII2009},  withÊ a  semi-numerical  scheme  to  describe  the  resulting ÊionizationÊ structure.  \cite{Kim2012a}  demonstrated  the  sensitivity  ofÊ the  ionization  structure  to  the  astrophysics  of  galaxy  formation, Êand  found  that  the  strength  of  supernovae  (SNe)Ê feedback  is  the  most  important  quantity.Ê

In addition to the 21~cm power spectrum, several studies have previously analysed the cross-power spectrum (correlation) between redshifted 21cm observations and galaxy surveys  \citep{furlanetto2007,lidz2009,lidz2011,Wiersma2013}. These models showed that the cross-power spectrum should be observable, but do not provide a self consistent link between the astrophysics of galaxy properties and the reionization structure. For example \cite{furlanetto2007} and \cite{lidz2009,lidz2011} used a simple one-to-one relation between luminosity and dark matter halo mass. Conversely, in \cite{Wiersma2013}, the cross-power spectrum was predicted using a semi-numerical code for 21cm emission based on dark matter overdensity cross-correlated with a semi-analytic model for galaxies. As a result, the calculation did not include the direct relation between galaxies and ionization structure.  In this paper our aim is to determine whether the cross-power spectrum can be used to infer the properties of high redshift galaxy formation. We present predictions for the cross-power spectrum between 21cm emission and galaxies using the model of \citet[][]{Kim2012a} who directly combined detailed models of high redshift galaxy formation using GALFORM with a semi-numerical description, and predict the resulting redshifted 21cm power spectrum of different reionization histories. This model provides self-consistent results  because the ionizing sources and observed galaxies are the same. These galaxies include both the observed luminous galaxies and the low mass ($\sim 10^{8}{\rm M_{\odot}}$) galaxies thought to drive reionization.

 We begin in \S~\ref{model} and \S~\ref{scheme} by describing the implementation of GALFORM, our method for modelling the ionization structure, the cross-power spectrum and cross-correlation function, and the cross-correlation coefficient. The cross-power spectra from our method, and the effect of feedback processes on the cross-power spectra are presented in \S~\ref{results-cross}. In \S~\ref{detectability} we describe the observational uncertainty. We finish with some conclusions in \S~\ref{Summary}.
\section{The GALFORM galaxy formation model}\label{model}
In this section we summarise the theoretical galaxy formation modelling based on \cite{Kim2012a} that is used in our analysis in order to describe the new features for this paper. 

We implement the GALFORM \citep{Cole2000} model, within the the Millennium-II dark matter simulation \citep{MII2009}. In this study, we specifically use the Lagos implementation of GALFORM  \citep{Lagos2012} model described in \citet[][]{Kim2012a}. The simulation has a cosmology including fractional mass and dark energy densities with values of $\Omega_{\rm m}=0.25$ , $\Omega_{\rm b}=0.045$ and $\Omega_{\Lambda}$=0.75, a dimensionless Hubble constant of $h$=0.73, and a power spectrum normalisation of $\sigma_{8}$=0.9. The particle mass of the simulation is 6.89$\times$10$^{6}$$h^{-1}{\rm M_{\odot}}$ and we detect haloes down to 20 particles (the minimum halo mass corresponds to $\sim 1.4\times 10^{8}h^{-1}{\rm M_{\odot}}$) in the simulation box of side length $L=100h^{-1}$Mpc.

\begin{figure}
\begin{center}
\includegraphics[width=8.6cm]{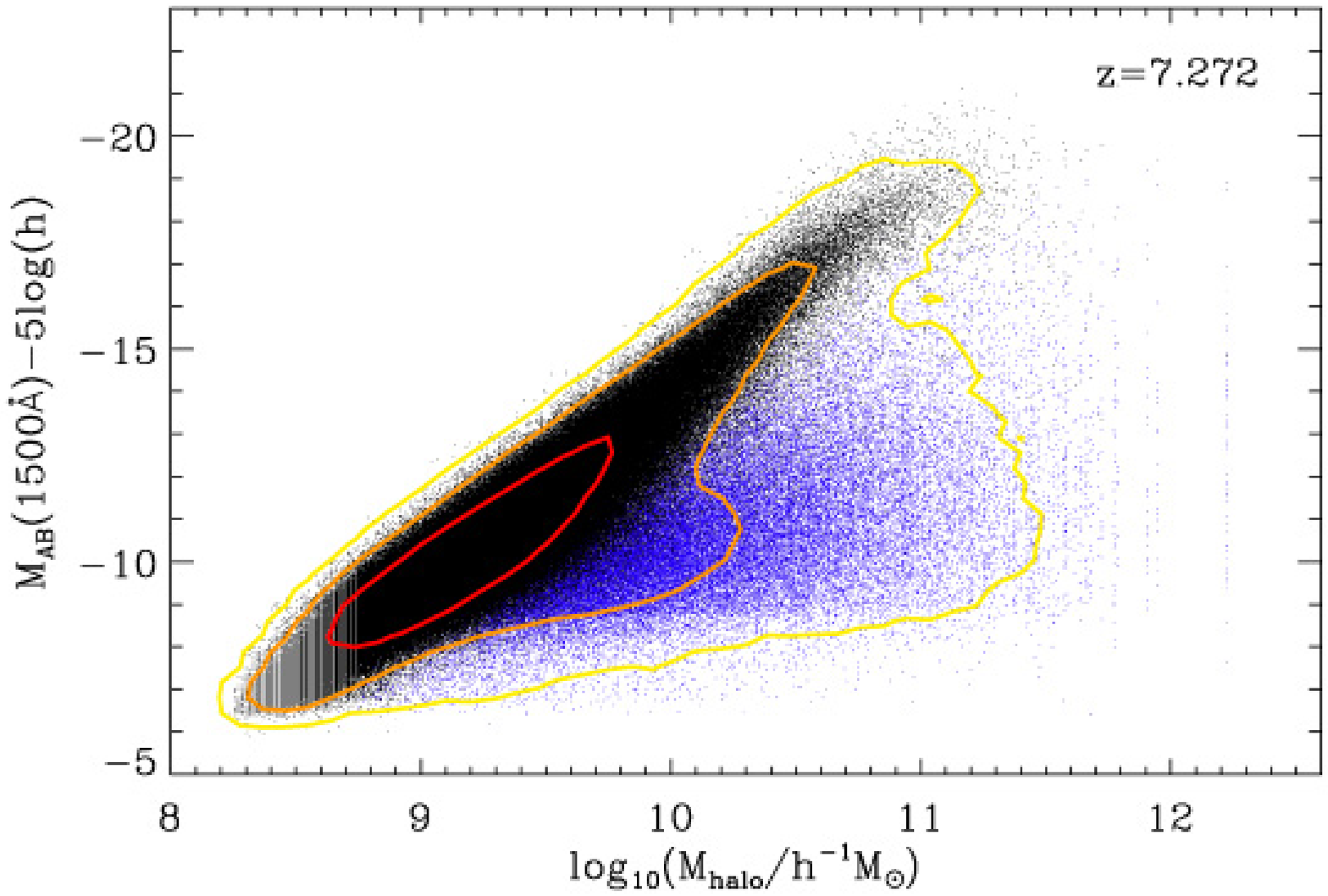}
\includegraphics[width=8.6cm]{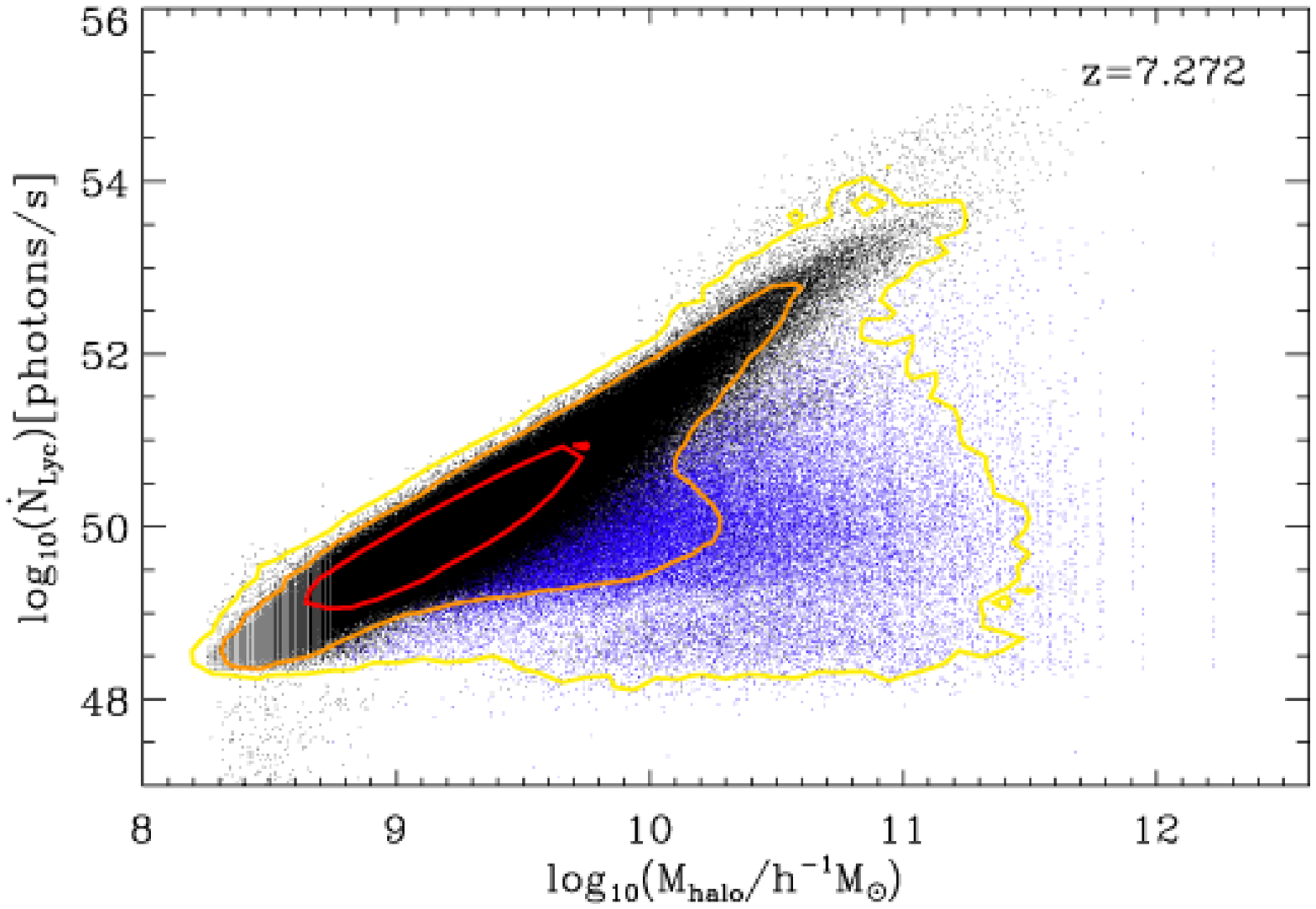}
\end{center}
\vspace{-3mm}
\caption{The relation between the UV magnitude (the rest-frame $1500{\rm \AA}$ AB magnitude) and the host halo mass (top), and between the total Lyman continuum luminosity ($\dot{N}_{\rm Lyc}$) of each galaxies (bottom) for galaxies and the host halo mass at $z=7.272$ from GALFORM. In each panel, black and blue dots represent central and satellite galaxies. Red, orange, and yellow colors represent  1 ($68.3\%$), 2 ($95.4\%$) and 3-sigma ($99.7\%$) levels. 
}
\label{distribution}
\end{figure}

Figure~\ref{distribution} shows the relation between the UV magnitude (the rest-frame $1500{\rm \AA}$ AB magnitude) including the effects of dust extinction of galaxies and the host halo mass (top), and between the total Lyman continuum luminosity ($\dot{N}_{\rm Lyc}$) of each galaxies and the host halo mass (bottom) from the GALFORM model. 
Of particular note is that the luminosity of an ionizing source is not simply proportional to the host halo mass as is often assumed in reionization models \citep{Iliev2011,lidz2009,lidz2011}. In part this is because of the distribution of satellite galaxies. The broad scatter of the relation indicates that physically motivated modelling for ionizing sources during the reionization should be included to understand the epoch of reionization. We note that this magnitude is not the same as ionizing luminosity. However, as shown in Figure~\ref{distribution},  the UV magnitude (the rest-frame $1500{\rm \AA}$ AB magnitude) is closely related to the ionizing luminosity. Furthermore, Figure~\ref{distribution} shows that the predicted ionizing luminosity to mass ratio from the model is not a simple one-to-one relation between luminosity and dark matter halo mass.

\section{The ionization model}
\label{scheme}
In this section we summarize the calculation of the ionized structure (\S~\ref{schemeSN}) and describe calculation of cross-power spectrum and cross-correlation function (\S~\ref{cross-pk} and \S~\ref{cross-xi}). 
\begin{figure*}
\begin{center}
\includegraphics[width=8.6cm]{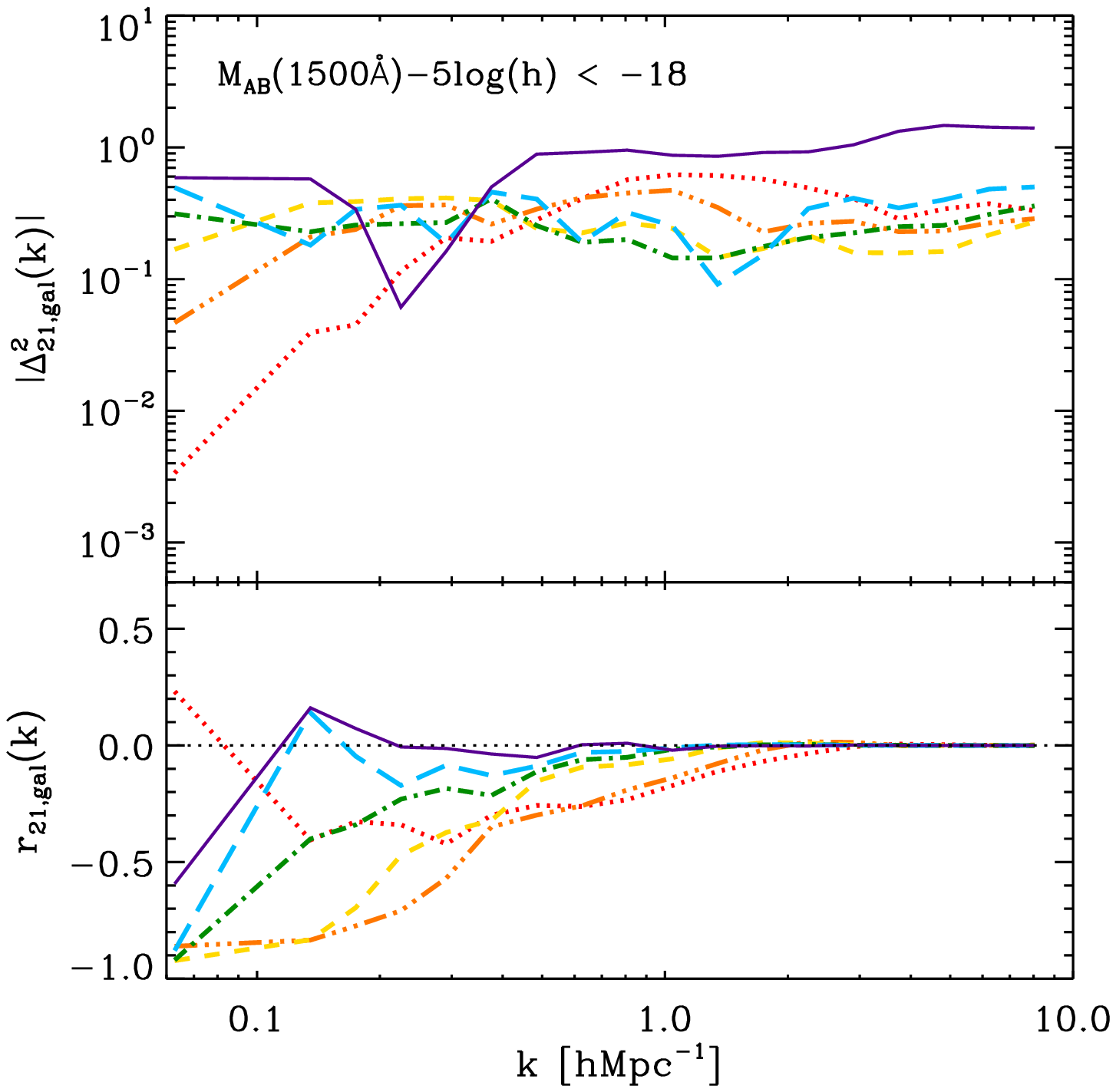}
\includegraphics[width=8.6cm]{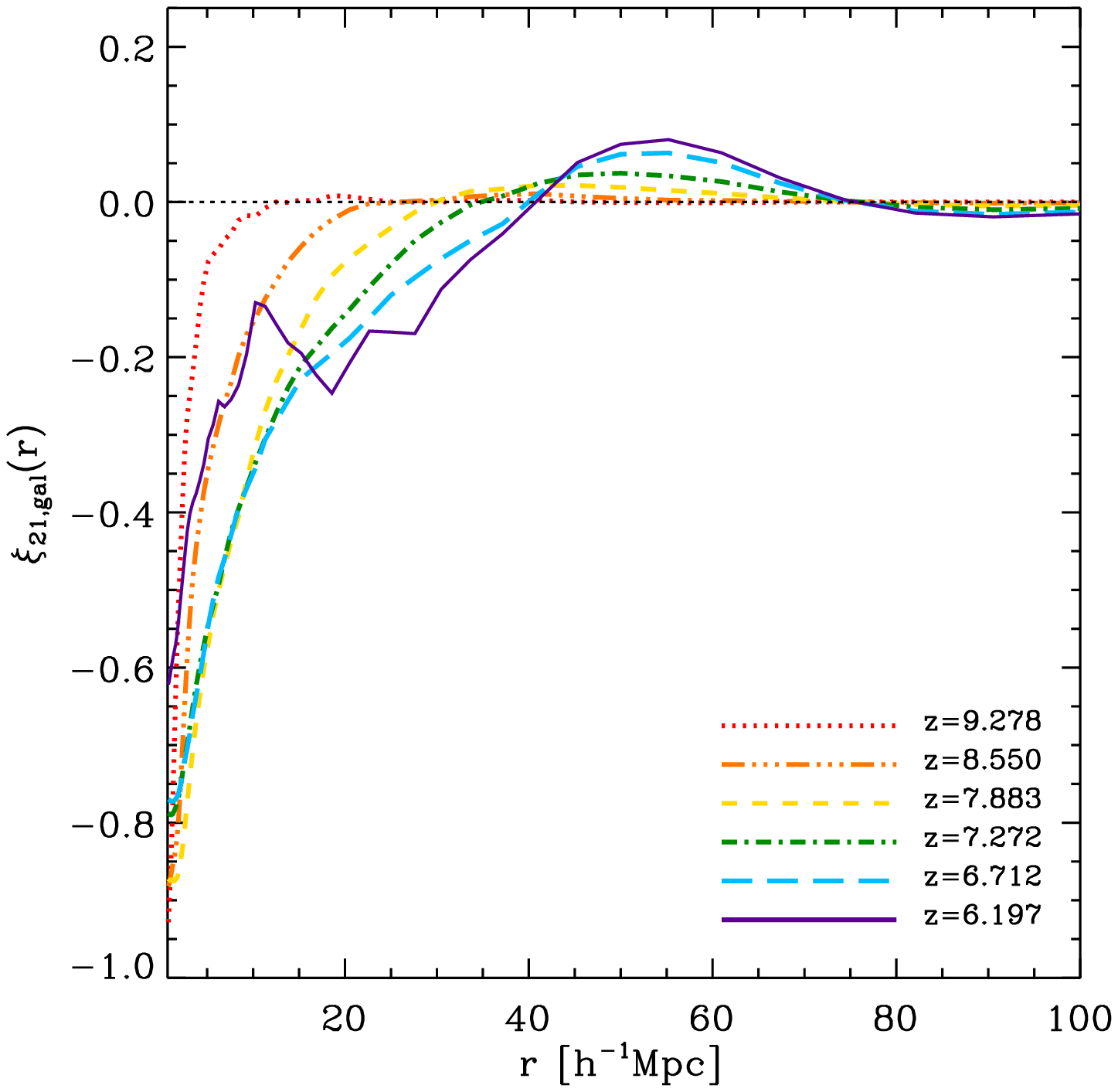}
\end{center}
\vspace{-3mm}
\caption{Redshift evolution of the cross-power spectrum and cross-correlation function between 21cm fluctuations and the galaxies which have the UV magnitude less than -18 in the model. Left panel: the absolute value of the cross-power spectrum (top) and cross-correlation coefficient (bottom). Right panel: The corresponding cross-correlation function. In each panel, dotted (red), dash-three dotted (orange), dashed (yellow), dash-dotted (green), long-dashed (blue), and solid (purple) lines represent results from at $z$ ($\left< x_{i} \right>$)~=~9.278 (0.056), 8.550 (0.16), 7.883 (0.36), 7.272 (0.55), 6.712 (0.75), and 6.197(0.95), respectively.
                } 
\label{pk_magUA18}
\end{figure*}

\subsection{Semi-Numerical scheme to calculate the evolution of ionizationed structure}
\label{schemeSN}

\citet{MF07} introduced an approximate but efficient method for simulating the reionization process, referred to as a semi-numerical technique.  In this paper we apply a semi-numerical technique to find the ionization structure resulting from GALFORM galaxies within the Millennium-II dark matter simulation. 

The simulation box is divided into cells. We calculate the number of photons produced by galaxies in each cell that enter the IGM and participate in reionization to be
\begin{equation}\label{nphotons}
N_{\rm \gamma, cell}={\it f}_{\rm esc}\int^{t_{z}}_{0}\dot{N}_{\rm Lyc,cell}(t)~dt,
\end{equation}
where $f_{\rm esc}$ is the escape fraction of ionizing photons produced by galaxies. Here $\dot{N}_{\rm Lyc,cell}(t)$ is the total Lyman continuum luminosity of the $N_{\rm cell}$ galaxies within the cell expressed as the emission rate of ionizing photons { (i.e.  units of photons/s)}. 

The ionization fraction within each cell is calculated as  
\begin{equation}
\label{Qvalue}
Q_{\rm cell}=\left[{N_{\rm \gamma, cell} \over (1+F_{\rm c})N_{\rm HI, cell}}\right],
\end{equation}
where $F_{\rm c}$ denotes the mean number of recombinations per hydrogen atom and $N_{\rm HI, cell}$ is the number of neutral hydrogen atoms within a cell. We assume that the overdensity of neutral hydrogen follows the dark matter and self-reionization of a cell occurs when $Q_{\rm cell}\geq1$. It is complicated to theoretically predict the values of $F_{\rm c}$ and ${\it f}_{\rm esc}$, and the values are not known. In this paper, we use the values of $(1+F_{\rm c})/f_{\rm esc}$ in table 2 of \cite{Kim2012a}. These parameters provide a reionization history with a mass-averaged ionization fraction of $\left< x_{i} \right>=0.55$ at $z=7.272$ and $\left< x_{i} \right>=0.75$ at $z=6.712$. We divide the Millennium-II simulation box into 256$^{3}$ cells, yielding cell side lengths of 0.3906$h^{-1}$Mpc and comoving volumes of 0.0596$h^{-3}$Mpc$^{3}$.

Based on equation~(\ref{Qvalue}), individual cells can have $Q_{\rm cell}\geq1$. On the other hand, cells with $Q_{\rm cell}<1$ may be ionized by photons produced in a neighbouring cell. In order to find the extent of ionized regions we therefore filter the $Q_{\rm cell}$ field using a sequence of real space top hat filters of radius $R$ (with $0.3906<R<100h^{-1}$Mpc), producing one smoothed ionization field $Q_R$ per radius. At each point in the simulation box we find the largest $R$ for which the filtered ionization field is greater than unity (i.e. ionized with $Q_R\geq1$). All points within the radius $R$ around this point are considered ionized. Ionization cells with $0<Q_{\rm cell}<1$ which are not part of an ionized $Q_{R}\geq1$ region retain their values. 

\subsection{The cross-power spectrum}\label{cross-pk}
The 21cm brightness temperature contrast may be written as
\begin{equation}\label{21cm_temp}
\tilde{\delta}_{\rm{21}}(\mathbf{r}) = T_{0}(z)[1-Q(\mathbf{r})] (1+\delta_{\textrm{DM,cell}}),
\end{equation}
where $T_{0}(z)=23.8\left( \frac{\Omega_{\rm b}h^2}{0.021}\right)\left[\left( \frac{0.15}{\Omega_{\rm m}h^2}\right) \left( \frac{1+z}{10} \right)\right]^{\frac{1}{2}}~{\rm mK} $ \citep{ZFH04}. For convenience, we define $\delta_{\rm{21}}(\mathbf{r})\equiv \tilde{\delta}_{\rm{21}}(\mathbf{r})/T_{0}(z)$, so that $\delta_{\rm{21}}(\mathbf{r})$ is a dimensionless quantity.
 Galaxy overdensity is given by
\begin{equation}\label{over_density}
\delta_{\rm{gal}}(\mathbf{r}) = \frac{\rho_{\rm{gal}}(\mathbf{r})-\bar{\rho}_{\rm{gal}}}{\bar{\rho}_{\rm{gal}}},
\end{equation}
where $\rho_{\rm{gal}}(\mathbf{r})$ is a galaxy density field and $\bar{\rho}_{\rm{gal}}$ is mean density.  Defining  $\hat{\delta}_{\rm{21}}(k)$ to be the Fourier transform of $\delta_{\rm{21}}(k)$, the cross-power spectrum is given by
\begin{equation}\label{cross_pk}
\left< \hat{\delta}_{\rm{21}}(\mathbf{k}_{\rm 1})\hat{\delta}_{\rm{gal}}(\mathbf{k}_{\rm 2}) \right> \equiv (2\pi)^3\delta_{D}(\mathbf{k}_{\rm 1}+\mathbf{k}_{\rm 2})P_{\rm{21,gal}}(\mathbf{k}_{\rm 1}),
\end{equation}
where $\delta_D(k)$ is the Dirac delta function. The dimensionless cross-power spectrum is
\begin{equation}\label{dless_pk}
\Delta^{2}_{\rm 21,gal}(k)=\frac{k^{3}}{(2\pi^{2})}P_{\rm 21,gal}(k).
\end{equation}

\subsection{The cross-correlation function}\label{cross-xi}
The cross-correlation function is defined as
\begin{equation}\label{xi}
\xi_{\rm 1,2}(\mathbf{r}) = \left< \delta_{\rm 1}(\mathbf{x})\delta_{\rm 2}(\mathbf{x}+\mathbf{r}) \right>.
\end{equation}
We calculate the cross-correlation function using the Fourier transform, 
\begin{equation}\label{fourier_transfrom3}
\xi_{\rm 21,gal}(r) = \frac{1}{(2\pi)^3} \int P_{\rm 21,gal}(k)\frac{{\rm sin}kr}{kr}4\pi k^2\rm{d} \it{k}.
\end{equation}
We also calculate the cross-correlation coefficient,
\begin{equation}\label{rk}
r_{\rm 21,gal}(k) = \frac{P_{\rm 21,gal}(k)}{\sqrt{P_{\rm 21}(k)P_{\rm gal}(k)}}.         
\end{equation}
\begin{figure*}
\begin{center}
\includegraphics[width=8.6cm]{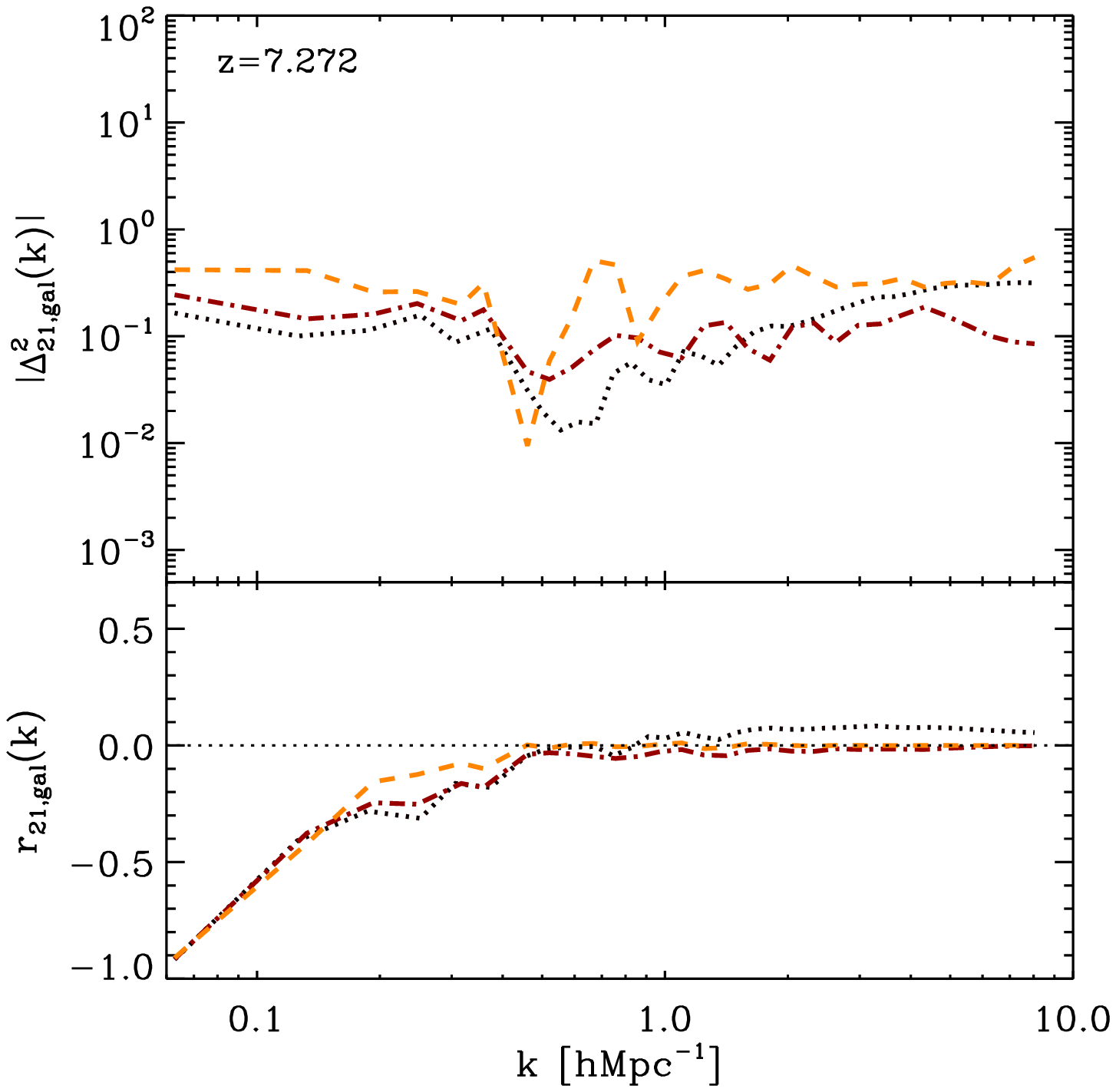}
\includegraphics[width=8.6cm]{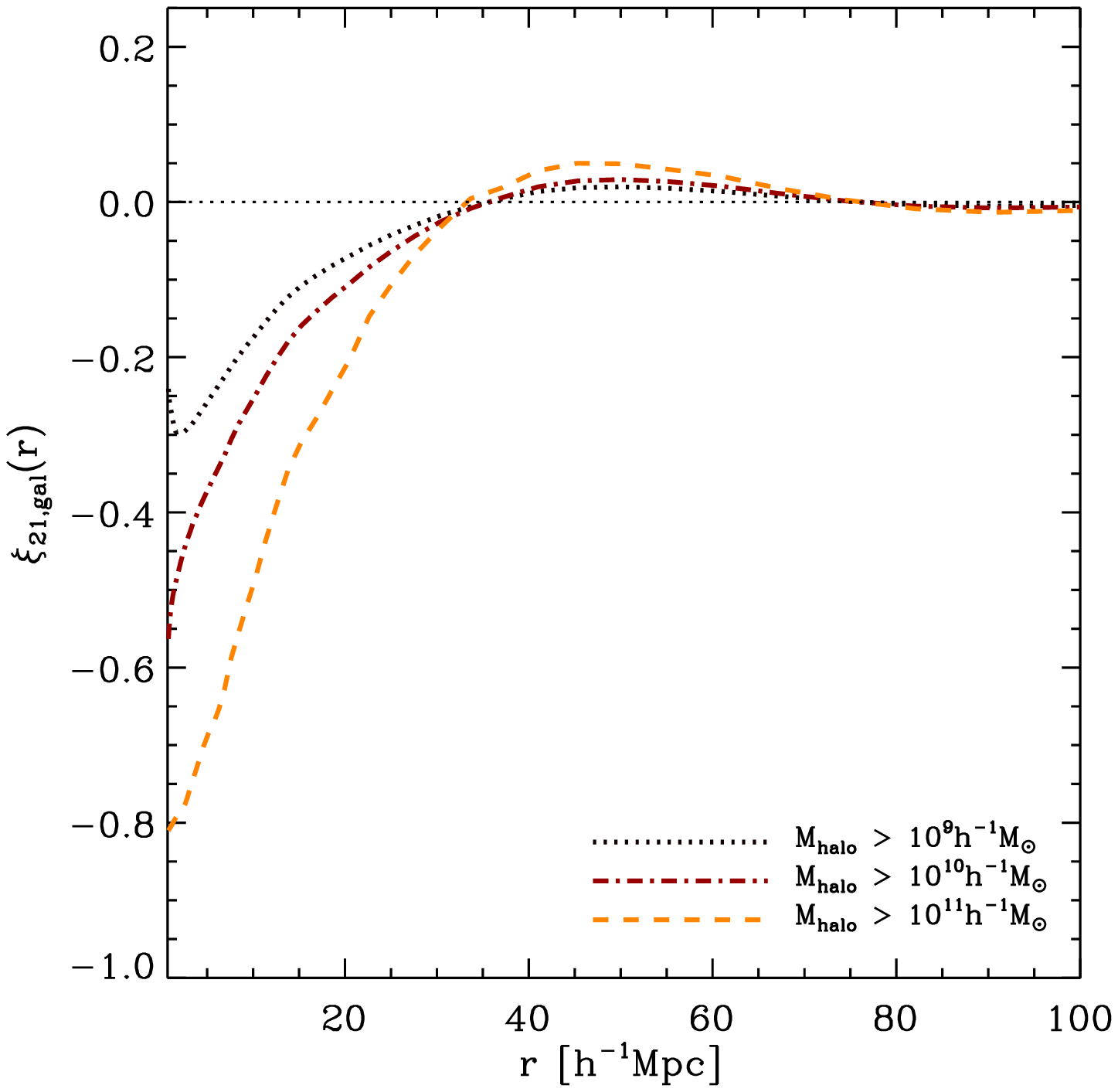}
\end{center}
\vspace{-3mm}
\caption{Comparison of the cross-power spectrum and cross-correlation function for different host halo mass thresholds at $z=7.272$ ($\left < x_{i} \right >=0.55$).  Left panel: the absolute value of the cross-power spectrum (top) and cross-correlation coefficient (bottom). Right panel: the corresponding cross-correlation function. In each panel, the dotted (dark brown), dot-dashed (brown), dashed (orange), long-dashed (yellow) lines show the cross-correlation using galaxies which are included in $10^{9}$, $10^{10}$, $10^{11}$ and $10^{12}$ $h^{-1}{\rm M}_{\odot}$, respectively.} 
\label{pk_z_mhalo}
\end{figure*}

\section{The correlation between 21cm emission and galaxies}\label{results-cross}

In this section we present predictions for the cross-power spectrum, cross-correlation function and cross-correlation coefficient  between 21cm emission and galaxies as a function of redshift, luminosity, and host halo mass (\S~\ref{prediction_cross}). We also discuss the effect of feedback processes on the cross-power spectrum, cross-correlation function and cross-correlation coefficient (\S~\ref{feedback_effect}).

\subsection{Predictions for the correlation between 21cm emission and galaxies}\label{prediction_cross}
Figure~\ref{pk_magUA18} shows the redshift evolution of the cross-power spectrum (top-left) and cross-correlation coefficient (bottom-left panel), and of the cross-correlation function (right panel) between redshifted 21cm emission and galaxies. We show three examples which have UV magnitude limits, $M_{\rm AB}(1500{\rm \AA})-5{\rm log}(h)<-18$, in the model. This magnitude threshold corresponds to the deepest ``wide" area survey with Wide Field Camera 3/infrared and the Cosmic Assembly Near-Infrared Deep Extragalactic Legacy Survey on Hubble Space Telescope \citep{Bouwens2011b, Finkelstein2012}.  At each redshift, we calculate a mass-averaged ionization fraction, $\left < x_{i}\right>$. From the correlation function, galaxies and 21cm emission are anti-correlated at small separations while at large separations we find a weak correlation. These regions are separated by a transition wavenumber at which the cross-correlation coefficient and cross-correlation function change from negative to positive. Galaxies are correlated with 21cm emission on scales larger than the ionized regions, but anti-correlated on smaller scales. The size of ionization regions therefore corresponds to this transition wavenumber. We find that the transition wavenumber from negative to positive cross-correlation coefficient increases as redshift decreases since the size of ionized regions generated by galaxies increases as the Universe evolves.

\begin{figure*}
\begin{center}
\includegraphics[width=8.6cm]{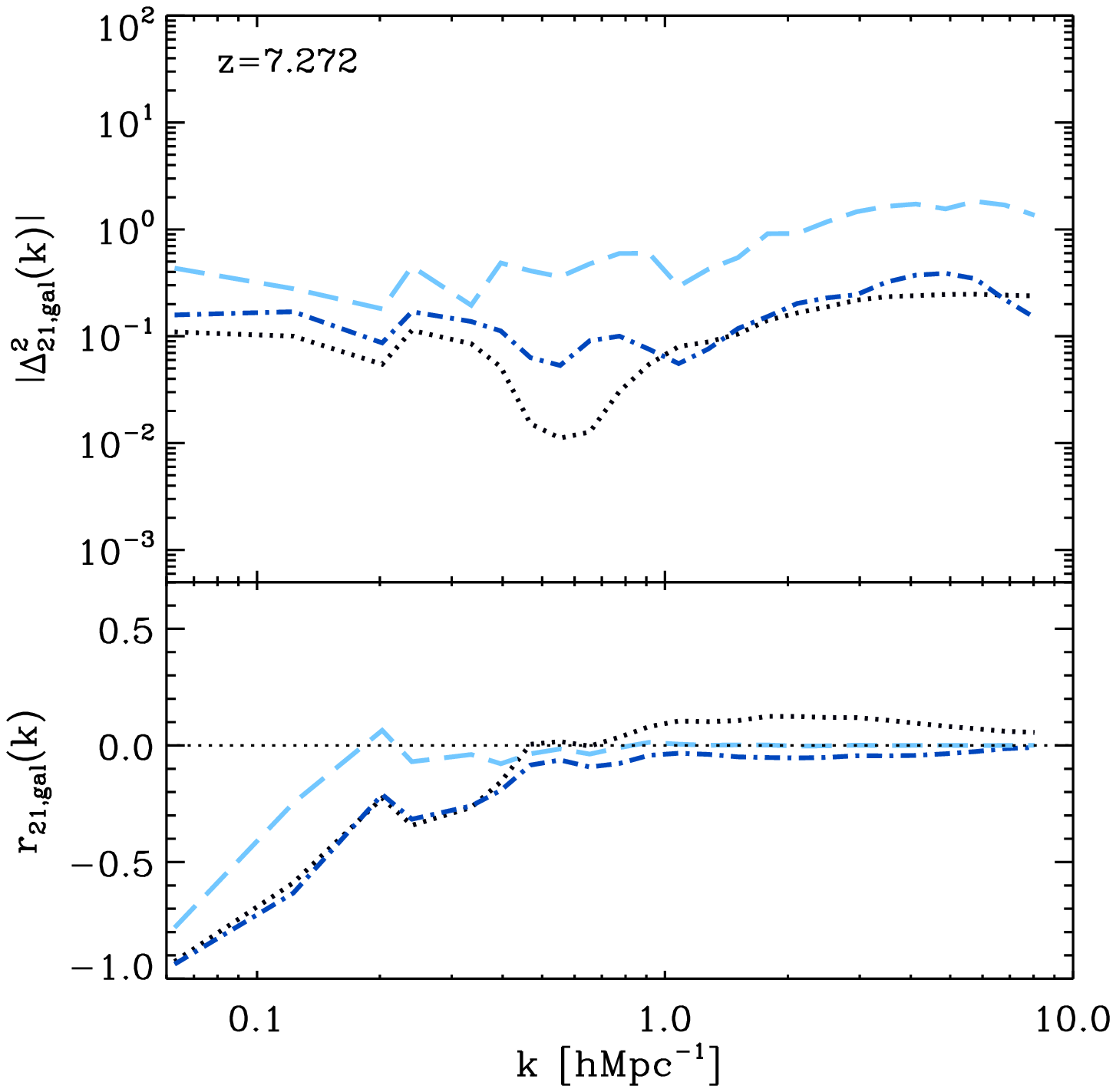}
\includegraphics[width=8.6cm]{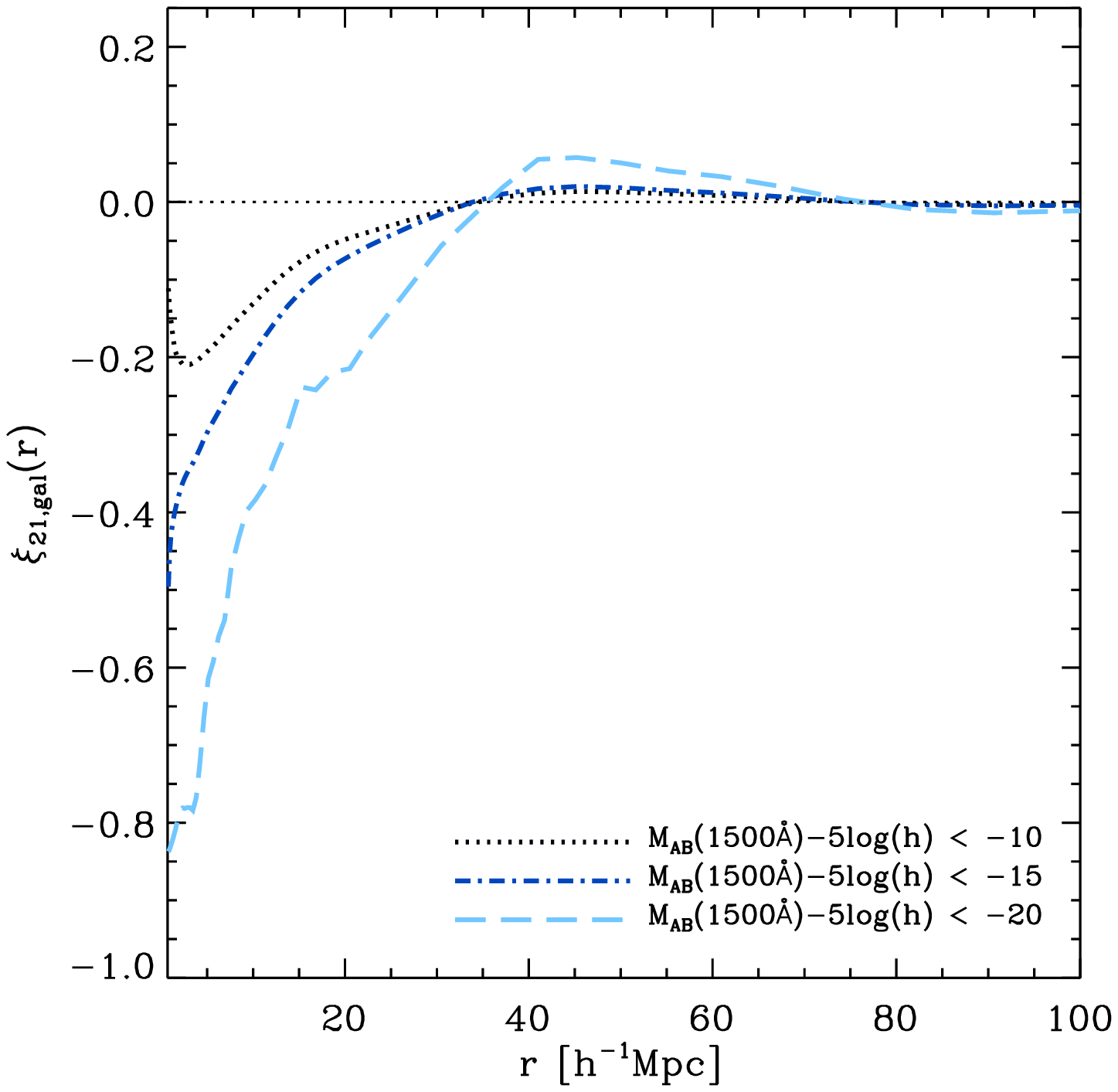}
\end{center}
\vspace{-3mm}
\caption{The same as figure \ref{pk_z_mhalo} but results are computed based on different UV magnitude  thresholds. In each panel, the dotted (black), dot-dashed (blue), and long dashed (sky-blue) lines show the cross-correlation using galaxy samples with are, respectively, more luminous than magnitude limits of -10, -15, and -20.
                } 
\label{pk_z_magUA}
\end{figure*}
Figure~\ref{pk_z_mhalo} shows the comparison of cross-power spectra and cross-correlation functions between different host halo mass thresholds at $z=7.272$ ($\left < x_{i} \right >=0.55$). We find that more massive halos exhibit stronger anti-correlation as expected \citep{lidz2009,Wiersma2013}.  The same trend is also shown in Figure~\ref{pk_z_magUA} where we compare the results from calculations with different UV magnitude ($M_{\rm AB}(1500{\rm \AA})-5{\rm log}(h)$) thresholds. 
Figure~\ref{pk_z_magUA} shows that the transition wavenumber is similar for galaxy samples selected at different luminosity thresholds, since this scale is primarily set by the size of HII regions. 

\subsection{The effect of feedback processes}\label{feedback_effect}

In order to investigate the effect on the power spectrum of different feedback processes in galaxy formation, we follow a similar method to \cite{Kim2012a}. We use the \cite{Lagos2012} galaxy formation model as our fiducial case, and then consider two variants of this (hereafter called NOSN models) which have supernovae feedback turned off. We use two variants of the NOSN model. First, we consider the inclusion of photoionization feedback using $V_{\rm cut}=30~{\rm km/s}$, where $V_{\rm cut}$ is a threshold value of the host halo's circular velocity \citep{Kim2012a}. Second, we removed both supernovae feedback and photoionization feedback by setting $V_{\rm cut}=0~{\rm km/s}$. We refer to this second model as NOSN (no suppression) in this paper. Since turning off supernovae feedback in the \cite{Lagos2012} model changes the bright end of the UV luminosity function, we have changed some other parameters so that the NOSN models still match the observed UV luminosity functions at $z=7.272$. Specifically, we introduce a stellar initial mass function dominated by brown dwarfs, with $\Upsilon = 4$, and also reduce the star formation timescale in bursts by setting $f_{\rm dyn}=2$ and $\tau_{\rm \ast burst,min}=0.005$~Gyr (see \cite{Cole2000} and \cite{Lacey2011} for more details of these parameters).

\begin{figure*}
\begin{center}
\includegraphics[width=8.6cm]{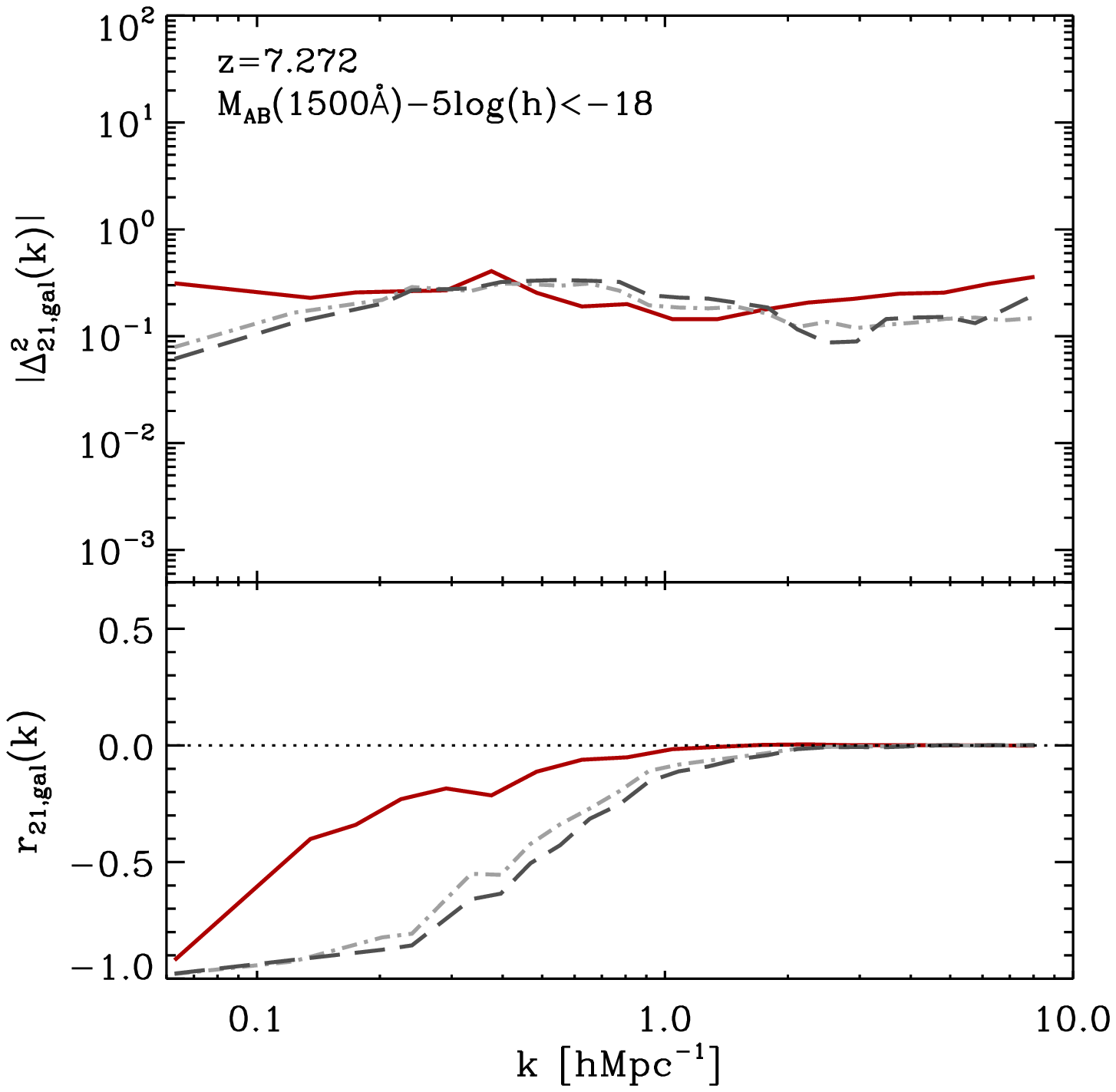}
\includegraphics[width=8.6cm]{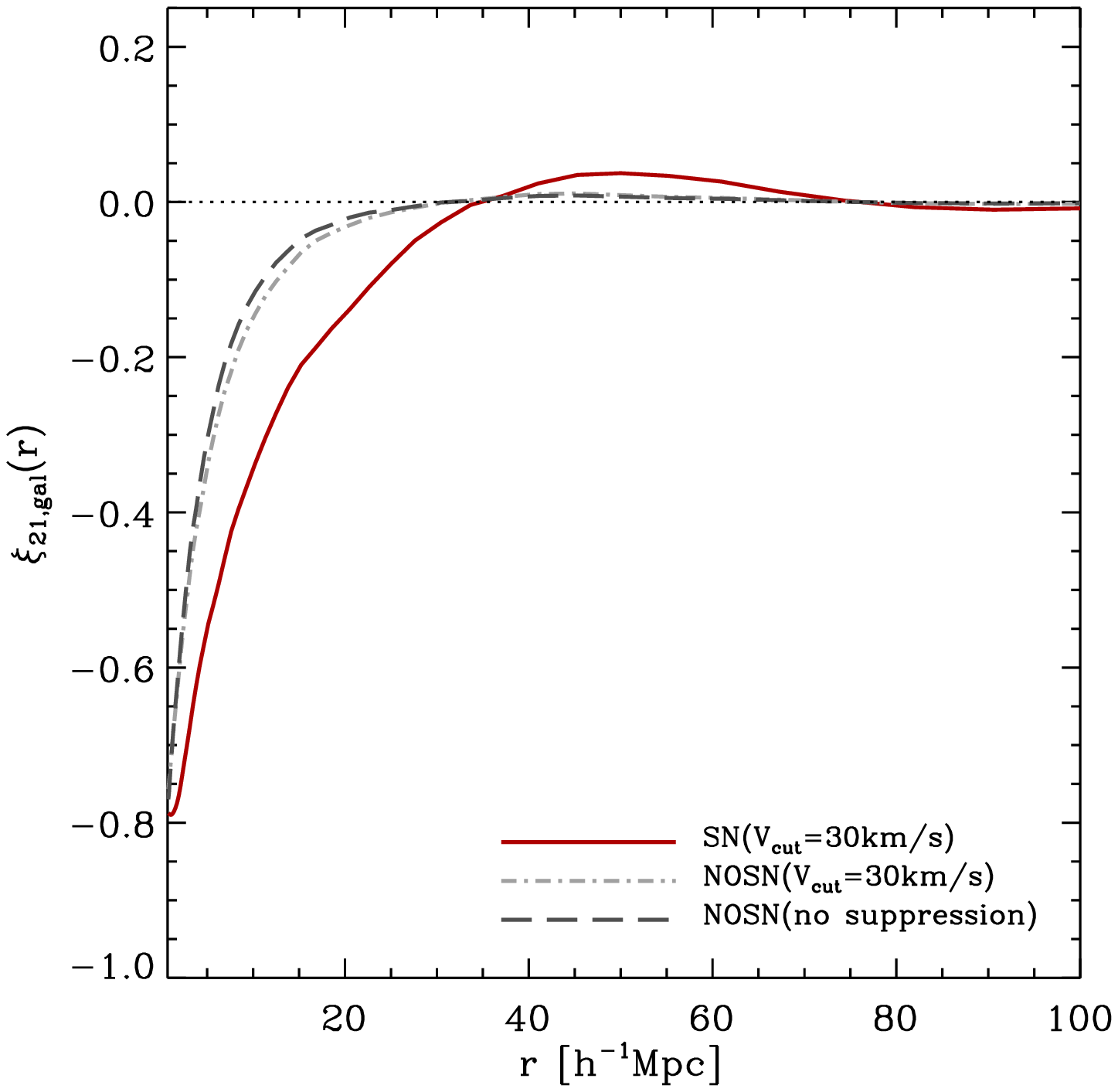}
\end{center}
\vspace{-3mm}
\caption{The same as Figure \ref{pk_z_magUA} but results are computed based on different feedback processes. In each panel, solid (red), dot-dashed (light grey), and long dashed (dark grey) lines represent our model, NOSN($V_{\rm cut}=30{\rm km/s}$), and NOSN(no suppression) models, respectively. 
} 
\label{pk_feedback}
\end{figure*}
In Figure~\ref{pk_feedback} we show the resulting comparison of cross-power spectra and cross-correlation functions at z=7.272. The locations of transition wave numbers between the \cite{Lagos2012} model and the two NOSN models are significantly different (see also ionization structure for these models in \cite{Kim2012a}). In particular, the \cite{Lagos2012} model has a larger transition scale.  On small scales, the cross-correlation function of the \cite{Lagos2012} model shows stronger anti-correlation than the two NOSN models between 21cm emissions and  galaxies.  Furthermore, we find that while photoionization feedback also suppresses low luminosity galaxies, the effect is smaller than the effect of the supernovae feedback.

\section{Detectability}\label{detectability}
In this section we describe the error estimation of the cross-correlation coefficient (\S~\ref{error_estimate}) and discuss observational requirements for future galaxy surveys (\S~\ref{results-error}). Our examples are based on the MWA-like observations of the 21cm signal combined with various hypothetical galaxy redshift surveys.

\subsection{Error estimate in the cross-correlation coefficient}\label{error_estimate}
In order to estimate the sensitivity of future surveys, we calculate the error on the cross-correlation coefficient \citep{furlanetto2007,lidz2009}. For convenience we use the notation of \cite{lidz2009} for the cross-correlation coefficient, 
\begin{equation}\label{rk_notation}
r_{\rm 21,gal}(k) = \frac{P_{\rm{21,gal}}(k)}{\sqrt{P_{\rm{21}}(k)P_{\rm{gal}}(k)}} \equiv \frac{A(k)}{\sqrt{B(k)C(k)}}.
\end{equation}

The error on the cross-correlation coefficient can be written as
\begin{align}\label{error_rk}
\frac{\sigma_{r}^2}{r^2}(k) = {}&\frac{\sigma_{A}^2}{A^2}(k) + \frac{\sigma_{B}^2}{4B^2}(k) + \frac{\sigma_{C}^2}{4C^2}(k)  \nonumber
\\
{} &- \frac{\sigma_{AB}^2}{AB}(k) - \frac{\sigma_{AC}^2}{AC}(k) + \frac{\sigma_{BC}^2}{2BC}(k).\end{align}
This equation has variances of the cross-power spectrum between 21cm and galaxy, and the auto-power spectra of both the 21cm emission and galaxies. It also has the covariance between different pairs of power spectra. The components of equation~(\ref{error_rk}) are given by  
\begin{eqnarray}\label{error_A}
\sigma_{A}^2(k,\mu) & = & \rm{var}[P_{\rm{21,gal}}(k,\mu)] \nonumber \\
                                    & = & \frac{1}{2}\left[ P_{\rm{21,gal}}(k,\mu)+\sigma_{B}(k,\mu)\sigma_{C}(k,\mu)  \right],
\end{eqnarray}
\begin{eqnarray}\label{error_B}
\sigma_{B}^2(k,\mu) & = & \rm{var}[P_{\rm{21}}(k,\mu)] \nonumber \\
                                    & = & \left[ P_{\rm{21}}(k,\mu)+ \frac{T_{\rm{sys}}^2}{T_{0}^{2}}\frac{1}{B t_{\rm{int}}} 
                                               \frac{D^2\Delta D}{n(k_{\perp})} \left( \frac{\lambda^2}{A_e} \right)^2 \right]^2,
\end{eqnarray}
\begin{eqnarray}\label{error_C}
\sigma_{C}^2(k,\mu) & = & \rm{var}[P_{\rm{gal}}(k,\mu)] \nonumber \\
                                     & = & \left[ P_{\rm{gal}}(k,\mu) + n^{-1}_{\rm{gal}}e^{k^2_{\parallel}\sigma^2_{\chi}} \right]^2,
\end{eqnarray}
\begin{eqnarray}\label{error_AB}
\sigma_{AB}^2(k,\mu) & = & \rm{cov}[P_{\rm{21,gal}}(k,\mu),P_{\rm{21}}(k,\mu)] \nonumber \\
                                     & = & \left[ P_{\rm{21,gal}}(k,\mu) P_{\rm{21}}(k,\mu) \right],
\end{eqnarray}
\begin{eqnarray}\label{error_AC}
\sigma_{AC}^2(k,\mu) & = & \rm{cov}[P_{\rm{21,gal}}(k,\mu),P_{\rm{gal}}(k,\mu)] \nonumber \\
                                     & = & \left[ P_{\rm{21,gal}}(k,\mu) P_{\rm{gal}}(k,\mu) \right],
\end{eqnarray}
and
\begin{eqnarray}\label{error_BC}
\sigma_{BC}^2(k,\mu) & = & \rm{cov}[P_{\rm{21}}(k,\mu),P_{\rm{gal}}(k,\mu)] \nonumber \\
                                     & = & \left[ P_{\rm{21}}(k,\mu) P_{\rm{gal}}(k,\mu) \right],                             
\end{eqnarray}
where $\mu$ is the cosine of the angle between $\mathbf{k}$ and the line of sight.
To introduce large scale redshift space distortions we use the relation, $P(k,\mu)=(1+\beta\mu^2)^2P(k)$, where $\beta=\Omega^{0.6}_{m}(z)/b$ and $b$ is a bias factor, between the redshift space power spectrum and the real space \citep{Kaiser1987}. We use $b_{\rm gal}^2(k)=P_{\rm gal}(k)/P_{\rm DM}(k)$ which is scale dependent and assume $b_{\rm 21}=1$ for 21cm power spectrum.

The first term  in equation~(\ref{error_B}) comes from a sample variance within the finite volume of the survey and the second term comes from the thermal noise of the 21cm telescope.  We have assumed specifications of the MWA for the calculation of thermal noise. In the thermal noise term, $T_{\rm sys} \sim 250[(1+z)/7]^{2.6} \rm K$ denotes the system temperature of the telescope; $B = 8{\rm MHz}$ is the survey bandpass; $t_{\rm int}$ is the integration observing time. We use 1000 hours total observing time in this calculation; $D$ and $\Delta D$ are the comoving distance to the survey volume and the comoving survey depth, $\Delta D=1.7\left(\frac{B}{0.1{\rm MHz}}\right)\sqrt{\frac{1+z}{10}}\left(\frac{\Omega_{\rm m}h^2}{0.15}\right)^{-1/2}$ \citep{FOB06}, respectively; $n(k_{\perp})$ denotes the number density of baselines in observing the transverse component of the wave vector, where $k_{\perp}=\sqrt{1-\mu^2}k$. Observing the signal of $k_{\perp}$ in each Fourier cell is related to the length of baseline and the antenna configuration. Here, we follow the method of \cite{Morales2005}, \cite{Bowman_etal2006} and \cite{Datta2007} for calculation of $n(k_{\perp})$. The maximum value of the transverse component of the wave vector is $k_{\perp,{\rm max}}=2\pi L_{\rm max}/(D\lambda)$, where $L_{\rm max}=750{\rm m}$ is the maximum baseline distance in the antenna array. This limit is due to the maximum angular resolution of the telescope related to $L_{\rm max}$. On the other hand, the minimum line-of-sight wavenumber is set by the bandpass $k_{\rm min}=2\pi/\Delta D$; The observed wavelength is $\lambda=0.21{\rm m}\times (1+z)$, and $A_{\rm e}$ is the effective collecting area of each antenna. We use  $A_{\rm e} \sim N_{\rm dip}\lambda^2/4$ \citep{Bowman_etal2006}, where $N_{\rm dip}=16$ is the number of dipoles. We have assumed 500 antennae elements.\footnote{The down scoped MWA has been constructed with 128 antennae. We use 500 here, corresponding to an upgraded array.}

From equations~(\ref{error_A}~--~\ref{error_BC}), we compute the errors of the power spectra averaged over a spherical shell of the logarithmic width $\epsilon = d{\rm ln}k$ for individual $k$-modes. For example, the error of the cross-power spectrum is given by
\begin{equation}\label{error_Ak}
\frac{1}{\sigma_{A}^{2}(k)} = \sum_{\mu} \frac{\epsilon k^3 V_{\rm{survey}}}{4\pi^2}
                                                   \frac{\Delta \mu}{\sigma^2_{A}(k,\mu)},
\end{equation}
where $V_{\rm survey}$ is the effective survey volume for a radio telescope, $V_{\rm survey}=D^2\Delta D (\lambda^{2}/A_{\rm e})$. The value of $\lambda^2/A_{\rm e}$ corresponds to the solid angle of the survey, which for the MWA corresponds to $\sim 800 {\rm deg^2}$. Note that if the galaxy survey volume is less than the 21cm survey volume, then the variance is increased by a factor of $V_{\rm survey,21}/V_{\rm survey,gal}$. The MWA is designed to operate at frequencies between 80 and 300MHz in order to observe the 21cm signal at $6<z<30$. When the 21cm signal is observed at $z\sim7$, the wavelength is $\sim 1.7$m corresponding to $\sim 200$MHz. 

\begin{figure}
\begin{center}
\includegraphics[width=8.6cm]{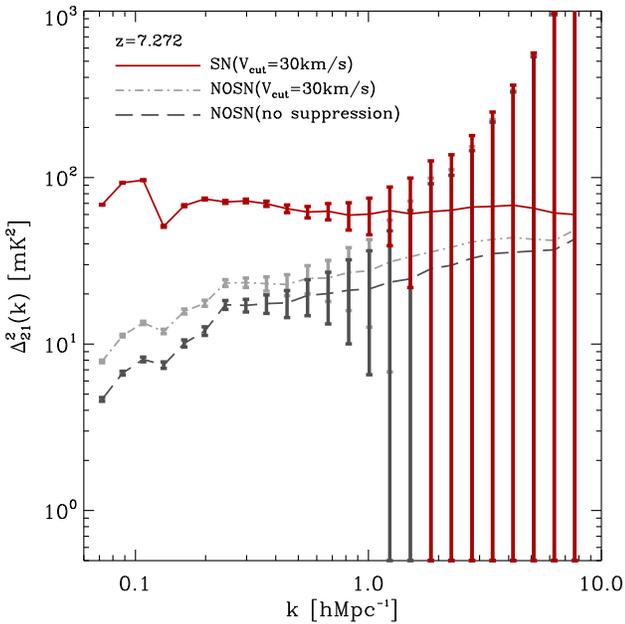}
\end{center}
\vspace{-3mm}
\caption{The 21cm power spectrum with estimated errors, based on an 800 $\rm deg^2$ survey area, at $z=7.272$. We assume 1000 hours total observing time, and, based on the assumption of 8MHz bandwidth, the survey depth is about 0.2 redshift units. Red represents the power spectrum from our model including supernovae feedback with $V_{\rm cut}=30{\rm km/s}$. The light  grey and dark grey lines represent power spectrum from the NOSN models with $V_{\rm cut}=30{\rm km/s}$ and no suppression, respectively. 
                }
\label{error21}
\end{figure}

Figure~\ref{error21} shows the 21cm power spectrum with errors estimated based on equation~(\ref{error_B})  for  cases including different feedback processes. The 21cm power spectra  show obvious differences between the models for supernovae feedback, especially at large scales. Figure~\ref{error21} reinforces the importance of detailed modeling of galaxy formation during  reionization \citep{Kim2012a}.

The error on the galaxy power spectrum is expressed in equation~(\ref{error_C}). The galaxy shot-noise is dependent on the number density of galaxies observable ($n_{\rm gal}$), $k_{\parallel} = \mu k$, and $\sigma_{\chi}=c\sigma_{z}/H(z)$, where $\sigma_{z}$ is the galaxy redshift error. Here, we assume a Gaussian distribution of redshift errors.  

\subsection{Observational requirements for future galaxy surveys}\label{results-error}

Following \cite{lidz2009}, we begin by considering a galaxy number density of $n_{\rm gal}=1.6\times10^{-4}~h^3{\rm Mpc^{-3}}$ for a survey in combination with 21cm observations from the MWA. To match this number density in our galaxy catalogue we use a magnitude  threshold,  with a value of  -19.4 at $z=7.272$ and -19.8 at $z=6.712$, in UV magnitude ($M_{\rm AB}(1500{\rm \AA})-5{\rm log}(h)$). We also match the number density of NOSN models in the same way. 

\begin{figure*}
\begin{center}
\includegraphics[width=5.8cm]{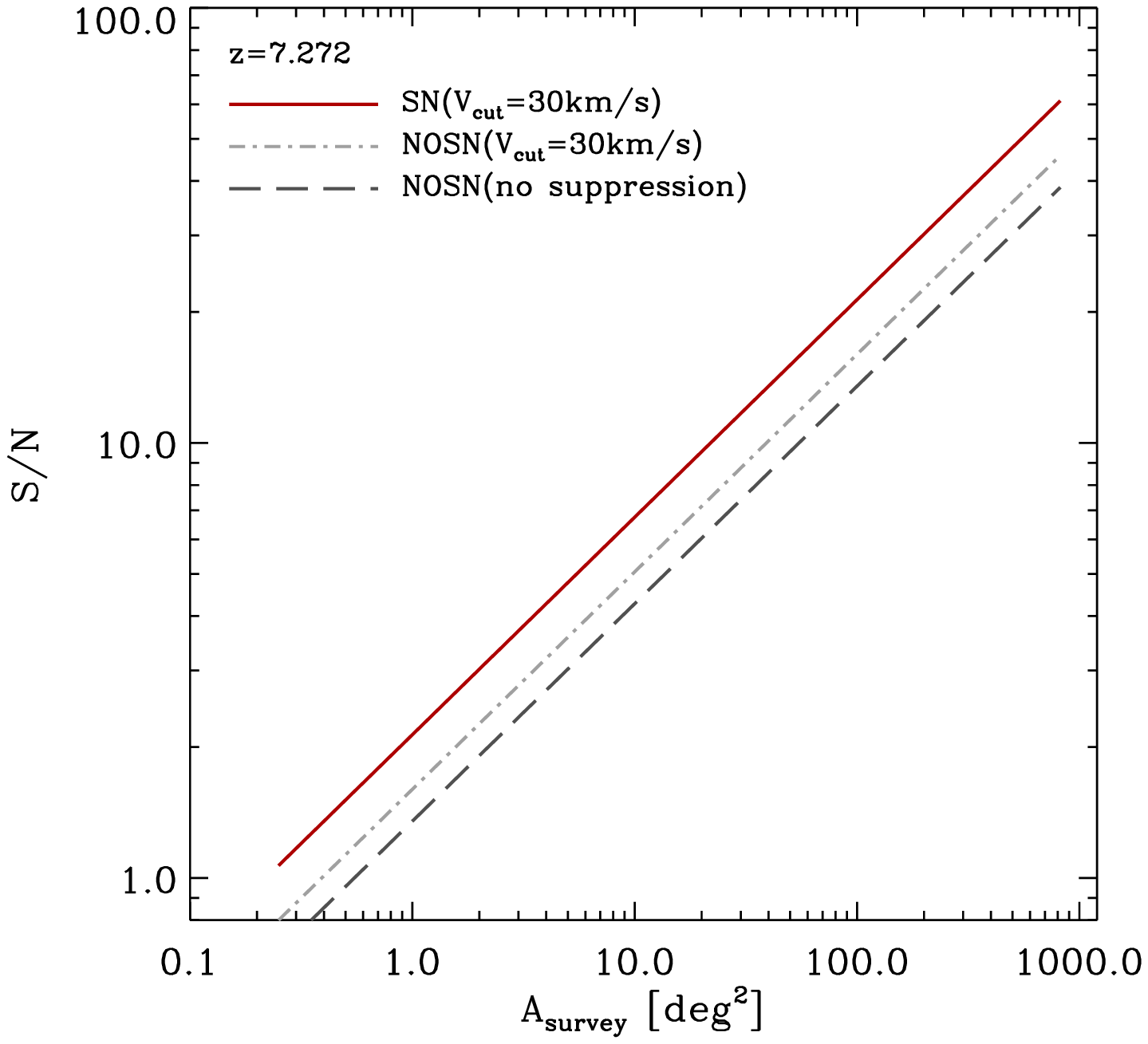}
\includegraphics[width=5.8cm]{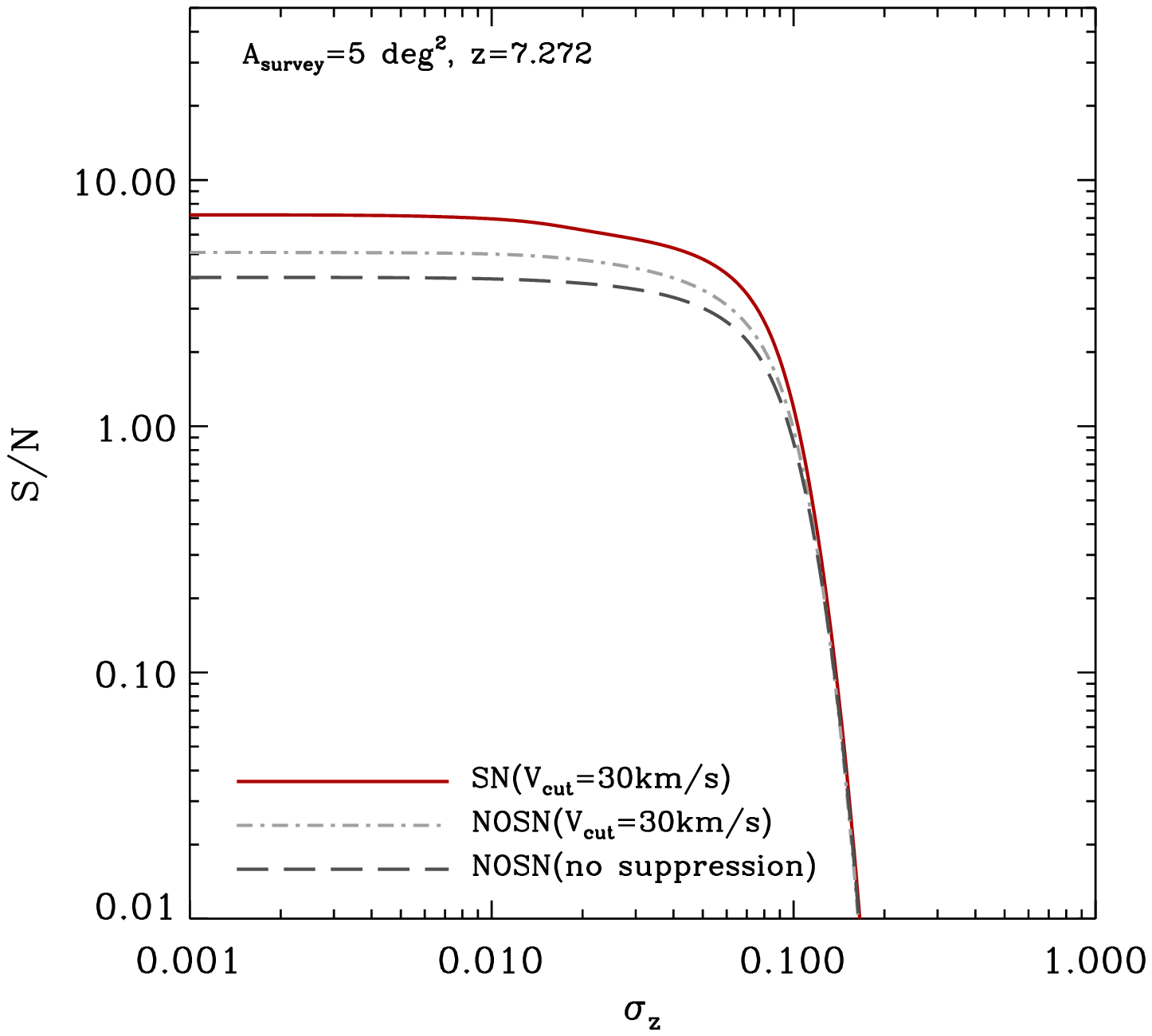}
\includegraphics[width=5.8cm]{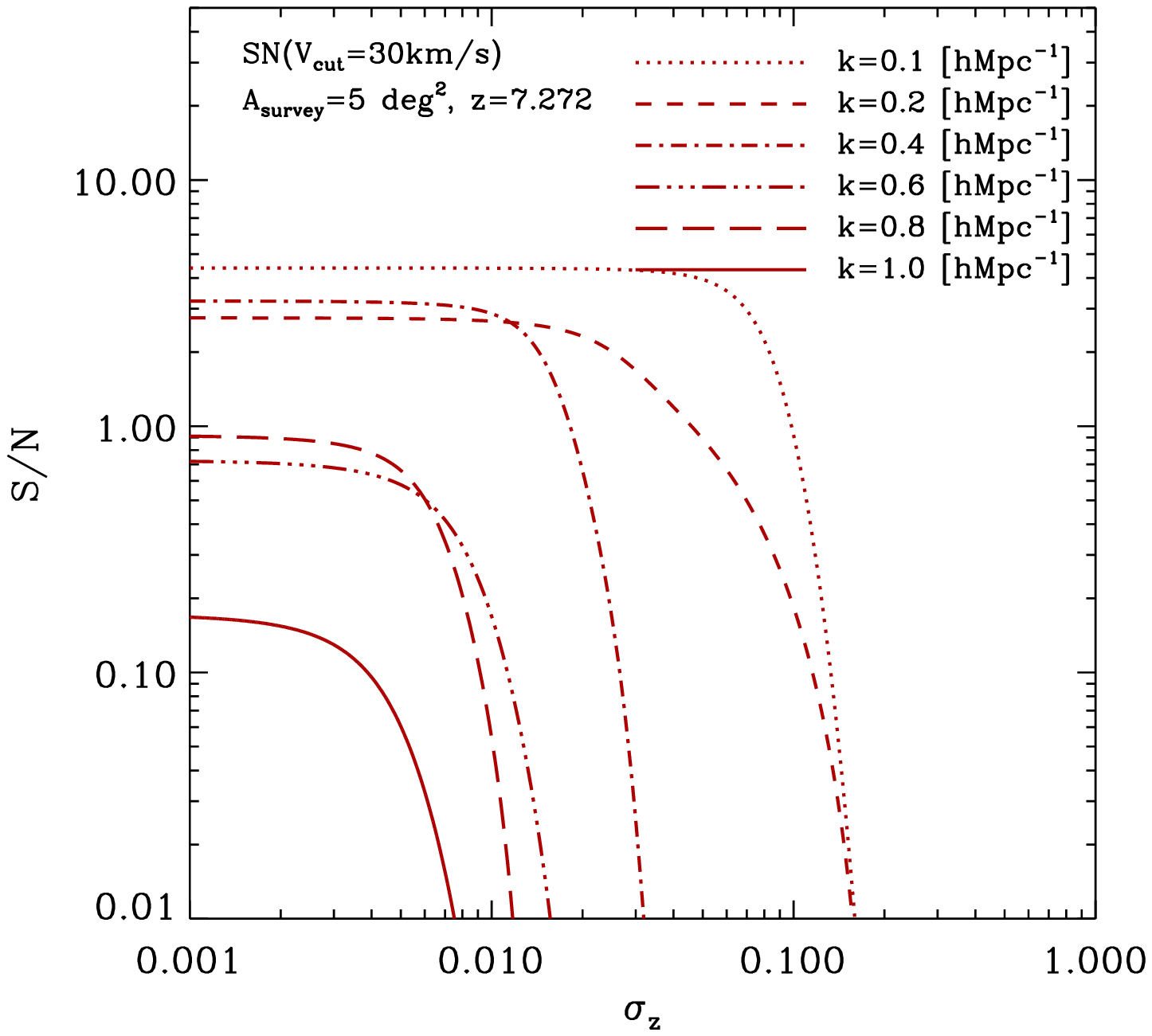}
\end{center}
\vspace{-3mm}
\caption{
The S/N for the cross-correlation coefficient as a function of survey area and relative redshift error at $z=7.272$. Left panel: plots of S/N as a function of survey area ($A_{\rm survey}$) for different models.  We assume $\sigma_{z}=0.05$. Central panel: a plot of S/N as a function of redshift error, $\sigma_{z}$, with $A_{\rm survey}=5~{\rm deg^2}$ for the default model. In the left and central panel, solid (brown), long-dashed (light grey), and dotted (dark grey) lines represent results from our model including SNe feedback with $V_{\rm cut}=30{\rm km/s}$ (the default model), the NOSN models with $V_{\rm cut}=30{\rm km/s}$ and no suppression, respectively. Right panel: plots of S/N as a function of $\sigma_{z}$ at different wave numbers for default model. Dotted, dashed, dot-dashed, three dot-dashed, long dashed and solid lines represent $k=0.1$, 0.2, 0.4, 0.6, 0.8, and 1.0 $h{\rm Mpc^{-1}}$, respectively. In each panel, we assume 1000 hours total observing time.}
               
\label{SNR}
\end{figure*}
To find the general requirements for detection of the cross-correlation, in Figure~\ref{SNR} we show the signal to noise (S/N) for the cross-correlation coefficient as a function of survey area ($A_{\rm survey}$) and redshift error ($\sigma_{z}$). In our calculations, we assume 1000 hours total observing time for the MWA. The left panel of Figure~\ref{SNR} shows the total S/N, which is calculated by summing up the S/N in each $k$ bin,
\begin{equation}\label{error_BC}
({\rm S/N})_{\rm total}^2 = \sum_i^{N_{\rm bin}} \left(\frac{\Delta k}{\epsilon k_{i}}\right)({\rm S/N})_{i}^{2},
\end{equation}
where $i$ represent $i$th bin and $\Delta k$ is the bin size. We assume a redshift error ($\sigma_{\rm z}$) of 0.05  as an example value from narrow-band survey for Lyman-$\alpha$ emitters \citep{Ouchi2008, Ouchi2010}. The S/N is calculated with a survey area of $800~{\rm deg^2}$, and then scaled the S/N with the relation, ${\rm S/N} \propto 1/\sqrt{A_{\rm survey}}$. Our default model that includes SNe feedback with $V_{\rm cut}=30{\rm km/s}$ shows increased S/N compared with the results of the NOSN models. The default model predicts a 3-$\sigma$ detection of cross-correlation with a $2~{\rm deg^2}$ survey area. Survey areas greater than $10~{\rm deg^2}$ will provide detailed high S/N measurements.

As a specific example, we also calculate the total S/N as a function of $\sigma_{\rm z}$ by assuming the survey area of $A_{\rm survey}=5~{\rm deg^2}$. The total S/N in the central panel of Figure~\ref{SNR} shows that measurements will require redshift uncertainties less than 0.1. The NOSN models show a similar shape to the default model, but have lower S/N. Lower accuracy redshifts ($\sigma_{\rm z}>0.1$) wash out the cross-correlation signal. An error of $\sigma_{z} \sim 0.1$ provides measurement only on larger scales ($k~[h{\rm Mpc^{-1}}]<0.2$) (right panel in Figure~\ref{SNR}). To measure the cross-correlation over a broad range of $k$, redshift uncertainties, $\sigma_{z}$, less than 0.01 will be required.

\begin{figure*}
\begin{center}
\includegraphics[width=8.6cm]{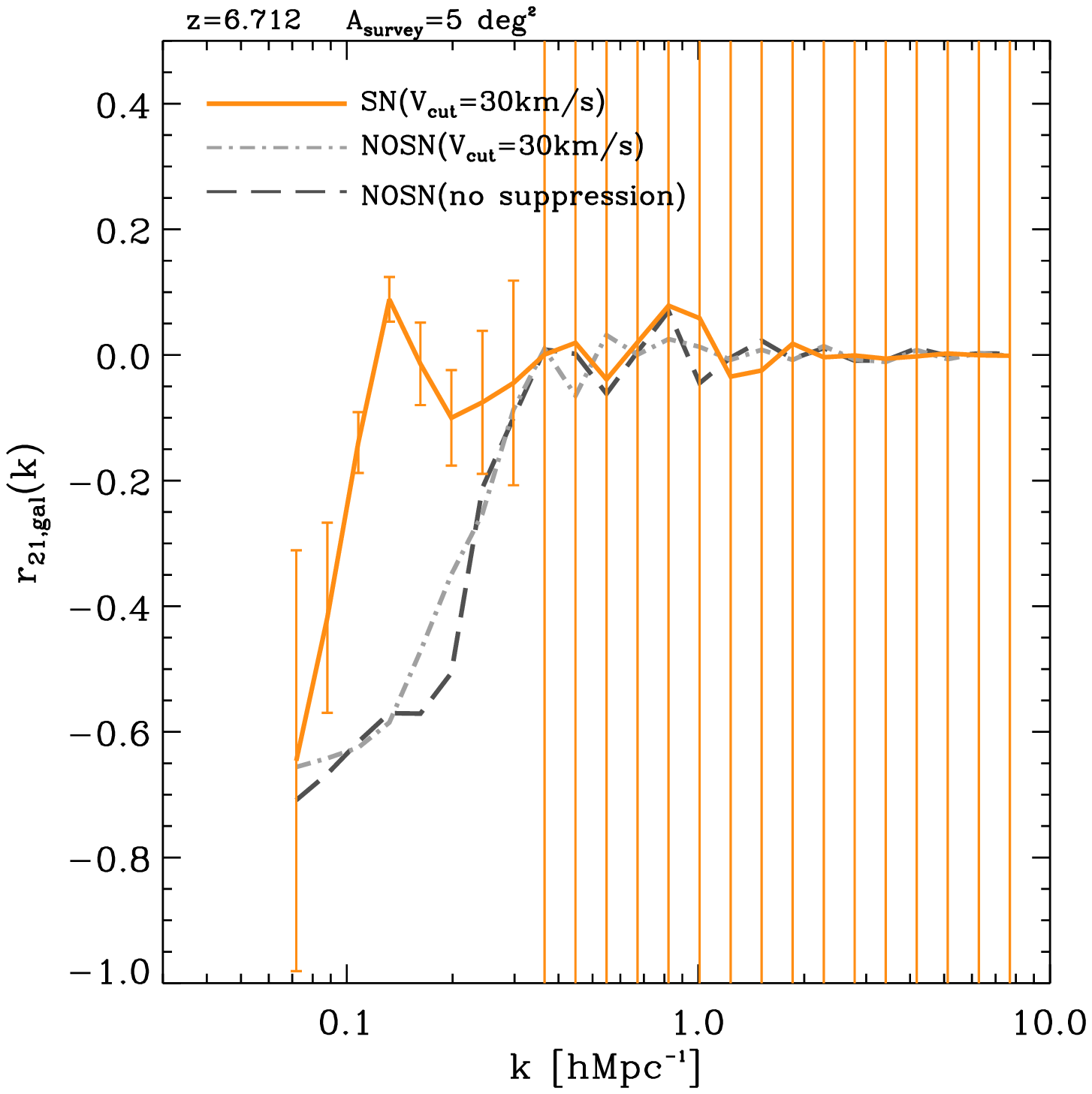}
\includegraphics[width=8.6cm]{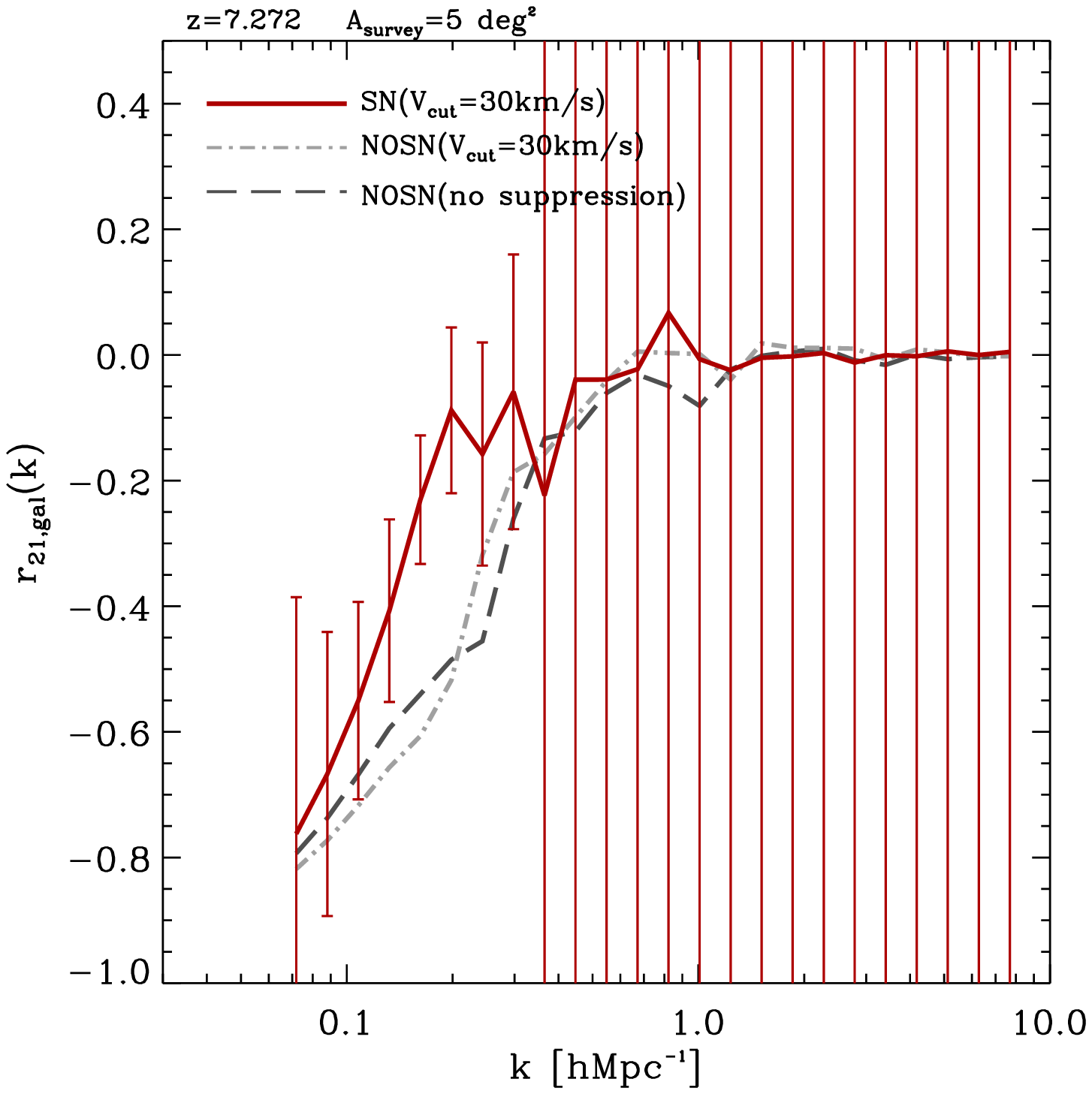}
\end{center}
\vspace{-3mm}
\caption{
The cross-correlation coefficient at $z=6.712$ (left panel) and 7.272 (right panel). The error bars are calculated for spherical bins of logarithmic width $\epsilon \equiv d{\rm ln}k =0.5$. We assume a $5~{\rm deg^2}$ galaxy survey field, 1000 hours total observing time, the redshift error of 0.05 and galaxy number density of the Subaru Deep Field survey. The solid(orange and red) lines represent the power spectrum from the default model \citep{Lagos2012} including supernovae feedback with $V_{\rm cut}=30{\rm km/s}$. The long dashed(light  grey) and dot-dashed (dark grey) lines represent power spectrum from the NOSN models with $V_{\rm cut}=30{\rm km/s}$ and no suppression, respectively. 
                } 
\label{error}
\end{figure*}

Figure~\ref{SNR} illustrates the conditions required for measurement of the 21cm-galaxy cross-correlation. Before concluding we discuss these requirements with respect to real galaxy surveys. In this study, we have used UV magnitude cuts to select galaxy samples which relate to observed Lyman-break galaxies. However, Lyman-break galaxies, which are photometrically selected, have $\sigma_{z} \gtrsim 0.5$ at $z\sim 6.5$ \citep{Beckwith2006} much longer than the $\sigma_{\rm z}<0.1$ requirement. As a result, Lyman-break surveys will not be sufficient to detect the cross-correlation \citep{Wiersma2013}. On the other hand, Ly-$\alpha$ emitters selected from narrow-band surveys have $\sigma_{z} \sim 0.05 - 0.1$ \citep{Ouchi2008, Ouchi2010}. Thus, a detection could be made based on the precision and volume of current Ly-$\alpha$ surveys. Our semi-numerical model does not predict Ly-$\alpha$ luminosity (see \cite{Orsi2008}). However, the difference between simulated populations of Ly-$\alpha$ emitters and Lyman-break galaxies is found not to be significant \citep{Dayal&Ferrara2012}. For the purpose of our calculation, we therefore use star forming galaxies with UV magnitudes ($M_{\rm AB}(1500{\rm \AA})-5{\rm log}(h)$) corresponding to the number density of  $n_{\rm gal}=1.6\times10^{-4}~h^3{\rm Mpc^{-3}}$, which is seen in the Subaru Deep Field at $z\sim 6.6$ \citep{Kashikawa2006}.

The largest Ly-$\alpha$ survey \citep{Ouchi2010} covered $1~{\rm deg^2}$ at $z\sim6.6$, and used a narrowband filter with a central wavelength of $9196~{\rm \AA}$ and a full width at half maximum of $132~{\rm \AA}$. These values correspond to a survey depth of $\Delta z\sim 0.11$ at $z=6.6$. This is smaller than, but comparable to the survey depth of MWA observation which is $\Delta z \sim 0.3$ corresponding to the bandwidth of $8~{\rm MHz}$ assumed for this paper. For the survey at $z\sim 7.3$, \cite{Shibuya2012} has a survey depth of  $\Delta z \sim 0.18$, which used the central wavelength of $10052~{\rm \AA}$ and a full width at half maximum of $214~{\rm \AA}$. This is also smaller than, but comparable to the MWA observation of the survey depth of $\Delta z\sim 0.38$. 

While $1~{\rm deg^2}$ represents the largest high redshift survey at the current time, future surveys will be larger. For example,  in the next 5 years Hyper Suprime-Cam on the Subaru telescope will observe $10^5$ galaxies at $z\sim5.7$ and 6.5 in a survey area of $\sim30~{\rm deg^2}$, and 100s of galaxies at $z\sim 7$ in a survey area of $3~{\rm deg^2}$ (M. Ouchi private communication). As shown in Figure~\ref{SNR}, this increased survey area will improve the S/N, so that the cross-power spectra signal could be detected with high significance. \footnote{The Subaru Deep Field is not accessible to the MWA. See \cite{Wiersma2013} for a calculation of LOFAR sensitivity.} 

Based on the requirement from \S~\ref{results-error}, we assume a $5~{\rm deg^2}$ galaxy survey field and a redshift error of 0.05 for a future galaxy survey.  We also assume 1000 hours total observing time. Figure~\ref{error} shows the predicted errors  for the cross-correlation coefficient within spherical bins of logarithmic width $\epsilon=0.5$ at $z=6.712$ and 7.272 for such a galaxy survey combined with the MWA.  The estimated errors are exponentially increased near the wave number of $1~h{\rm Mpc^{-1}}$, because of the limit of $k_{\rm \perp,max}$ for a 21cm survey. We compare the result with the cross-correlation coefficient from the NOSN models. The result shows that we could observationally distinguish our default model from two different NOSN models.  

Ly-$\alpha$ observations at $z \gtrsim 7$ over a large area are very challenging. The latest Ly-$\alpha$ survey at $z\sim 7.3$ \citep{Shibuya2012} has a galaxy number density of $\sim 6.7 \times 10^{-6}$ covering a survey area of  $~0.48~{\rm deg^2}$. This value is smaller than the value we assume for our error estimation. Computing the cross-power spectrum corresponding to this number density is not possible owing to the limited box size of our simulation. However, we have checked that the estimated error would approximately increase by a factor of 2, when using this number density. 



\section{Summary and conclusions}\label{Summary}
In this study we have investigated evolution of the cross-power spectrum, cross-correlation function and cross-correlation coefficient between 21cm emission and galaxies using the model of \citet[][]{Kim2012a}. This model combines the hierarchical galaxy formation model GALFORM  implemented within the Millennium-II dark matter simulation, with a semi-numerical scheme to describe the resulting ionization structure. We find that there is a  transition wave number, $k$, at which the cross-correlation coefficient changes from negative to positive \citep{lidz2009}. This transition wave number is associated with the size of the ionized regions generated by  galaxies, and increases with decreasing redshift. We also find the same trend in the cross-power spectrum and cross-correlation function. We calculated the cross-power spectrum as a function of UV luminosity ($M_{\rm AB}(1500{\rm \AA})-5{\rm log}(h)$) and host halo mass. These calculations reveal that bright galaxies and galaxies residing in massive halos have stronger anti-correlation, but a similar transition wavenumber.

We have studied  observational uncertainties in measurement of the cross-correlation coefficient based on the specifications of an upgraded (512 tile) Murchison Widefield Array (MWA) combined with galaxy surveys. The results show that the cross-power spectrum signal could be detected when combined with more than 3 square degrees of a galaxy survey at the depth of the future galaxy survey having redshift error $< ~ 0.1$. We have also investigated the dependance on the inclusion of feedback processes in the galaxy modelling. We find that the amplitude of the cross-correlation is larger when SNe feedback is considered and that the cross-correlation coefficient has a different shape compared to a model with no SNe feedback. Thus the cross-correlation could be used to determine the importance of SNe feedback in high redshift galaxies. Our results imply that detailed modelling of reionization processes and galaxy formation are required to predict an accurate cross-correlation between 21cm emission and galaxies, and to interpret future observational measurements

\vspace{5mm}

{\bf Acknowledgments} HSK is supported by a Super-Science Fellowship from the Australian Research Council. JSBW acknowledges the support of an Australian Research Council Laureate Fellowship. The Centre for All-sky
Astrophysics is an Australian Research Council Centre of Excellence, funded by grant CE110001020.  This work was supported in part by the Science and Technology Facilities
Council rolling grant ST/I001166/1 to the ICC. The Millennium II Simulation was
carried out by the Virgo Consortium at the supercomputer centre of the
Max Planck Society in Garching.
Calculations for this paper were partly performed on the ICC Cosmology Machine, which is part of the DiRAC Facility jointly funded by STFC, the Large Facilities Capital Fund of BIS,
and Durham University.  

\newcommand{\noopsort}[1]{}

\bibliographystyle{mn2e}

\bibliography{Cross}

\label{lastpage}
\end{document}